\documentclass{camera}
\usepackage{graphicx}  

\begin{document}

%
\title{Duration and hardness ratio of {\em Swift} GRBs}

%
\author{A. Gomboc \and D. Kopa\v c}

%
\organization{Faculty of Mathematics and Physics, University of Ljubljana, Slovenia}

\maketitle

\begin{abstract}
We review the $T_{90}$ duration and hardness ratio of {\em Swift} Gamma Ray Bursts (GRBs). We focus on GRBs with known redshift and review their gamma properties in the GRBs rest frames. We find that GRBs number vs. $T_{90}/(1+z)$ distribution shows a separation between two classes at 0.65~s. Furthermore, we find that the difference in hardness ratio between short and long bursts is not very pronounced and depends on energy channels used for comparison. 
\end{abstract}

%
\section{Introduction}
 GRBs detected by BATSE \cite{ref:mee} show a bimodality in the distribution of the duration (usually characterized with $T_{90}$)
\cite{ref:kou}. Classification in two classes, short ($T_{90}< 2~$s) and long ($T_{90}>2~$s) was also supported by the hardness ratio ($HR$)
showing the short bursts to be harder than long ones \cite{ref:kou}. This classification in short and long bursts is widely accepted, as well as the understanding that they represent two different physical phenomena. BATSE and {\em Swift} data also show some evidence for a third, intermediate group (\cite{ref:hor98}, 
\cite{ref:hor09}). For simplicity, we here classify bursts only to short and long.

Here we review the $T_{90}$ and hardness ratio of {\em Swift} GRBs detected between Dec 2004 and Dec 2009.
Our sample consists of 436 GRBs, among which there are 145 with known redshift $z$.
All data were taken from the {\em Swift} archive\footnotemark[1] 
and the Gamma ray bursts Coordinates Network.\footnotemark[2] 

\footnotetext[1]{http://swift.gsfc.nasa.gov/docs/swift/archive/grb$_-$table.html/}
\footnotetext[2]{http://gcn.gsfc.nasa.gov/swift$_-$gnd$_-$ana.html}
\section{GRBs duration in rest frame}

\begin{figure}[h]
\begin{center}
\includegraphics[angle=-90,scale=0.27]{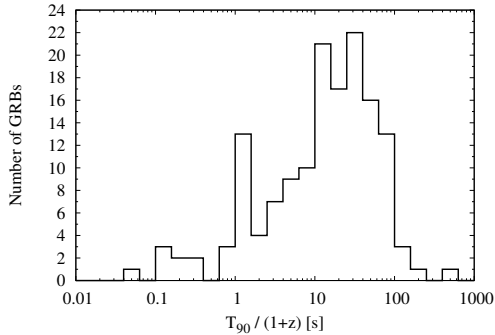}
\caption{$N_{\rm GRB}$ vs. $T_{90,\rm i}$. The separation between short and long bursts in the observer frame at $T_{90}=2~$s is shifted to $T_{90,\rm i} =0.65~$s in the rest frame.}
\label{fig01} 
\end{center}
\end{figure}

The ratio between the number of short and long bursts in the {\em Swift} sample is roughly 1:10, while in the BATSE sample it was about 1:3. The duration distribution of bursts in the First {\em Swift} BAT GRB catalogue does not show clear bimodality due to the smaller sample of short duration bursts \cite{ref:sak}. Our results on number of GRBs ($N_{\rm GRB}$) vs $T_{90}$ confirm this finding. 

To compare GRBs in their rest frames, we focus on GRBs with known redshift and take into account the cosmological time dilatation. Fig.~\ref{fig01} shows $N_{\rm GRB}$ vs. $T_{90,\rm i}=T_{90}/(1+z)$. 
There is a gap between short and long GRBs at $T_{l-s, i} =0.65~$s. This is consistent with the gap at $T_{l-s}=2~$s seen in the observer frame for BATSE bursts, but shifted to the 'average' rest frame, i.e. to the average redshift of bursts in our sample: $\bar{z}=2.1 (\pm 1.5, 1\sigma)$. We will use this gap in the final part of our analysis for classifying bursts into intrinsically short and long. (Taking into account the correlation between burst duration and energy \cite{ref:fen} would give $T_{90,\rm i}=T_{90}/(1+z)^{0.6}$,  $T_{l-s, i} =1.0~$s and same GRBs falling into the category of short/long bursts, therefore giving the same results as those presented below.) \\




\begin{table}
\begin{center}
\scriptsize
 \caption{Hardness ratios ($HR$s) of {\em Swift} GRBs for different energy channels (for details see text). }
 \label{HR}
 \begin{tabular}{l}

Observer frame (436 GRBs, $38$ short, $398$ long):
 \end{tabular}

\begin{tabular}{lcccccccc}
  \hline 
	&Short		&	&	& & Long & & &  \\
 & 		mean	 &$\sigma_{\rm mean}$ & median & $\sigma_{\rm med}$ &	mean &	$\sigma_{\rm mean}$  &median	 &$\sigma_{\rm med}$ \\
   \hline 
$HR_{4321}$ & 	2.46 &	1.46 &	2.09	 &1.04	 &	1.19 &	0.68 &	1.08 &	0.39 \\
$HR_{43}$	 &	4.29  &	1.92 &	3.90	 &1.4	 &	2.59 &	0.98 &	2.48 &	0.58 \\
$HR_{41}$	  &	22.49 &	21.27 &	14.26 &	9.48	 &	6.80 &	7.73$^*$ &	5.09 &	2.54 \\
$HR_{31}$	  &	4.23 &	2.32 &	3.65	 &1.65	 &	2.21 &	1.10 &	2.05 &	0.63 \\
$HR_{21}$	  &	2.22 &	0.61 &	2.16	 &0.48	 &	1.66 &	0.38 &	1.65 &	0.23 \\
$HR_{32}$	  &	1.76 &	0.54 & 	1.69	 &0.41	 &	1.26 &	0.33 &	1.25 &	0.20 \\
\hline
\end{tabular}

 \begin{tabular}{l}
Observer frame (145 GRBs with known $z$, $11$ short, $134$ long): \\
 \end{tabular}

\begin{tabular}{lcccccccc}
  \hline 
 & Short	& & & &				Long & & & \\
	 & 	mean	 &$\sigma_{\rm mean}$ &median &$\sigma_{\rm med}$ &	mean  &	$\sigma_{\rm mean}$ &median	 &$\sigma_{\rm med}$ \\
  \hline 
$HR_{4321}$ & 	1.92 &	1.19	 &1.67 &0.91	 &	1.19 &	0.59 &	1.05 &	0.41 \\
$HR_{43}$ &		3.59  &	1.60	 &3.33	 &1.36	 &	2.58 &	0.88 &	2.43 &	0.61 \\
$HR_{41}$	  &	14.96 &	14.08 &	10.0	 &7.13	 &	6.54 &	4.81 &	4.87 &	2.61 \\
$HR_{31}$	  &	3.39 &	1.9 &	3.01	 &1.47	 &	2.21 &	0.96 &	2.0 &	0.66 \\
$HR_{21}$	  &	2.01 &	0.54 &	1.97	 &0.49	 &	1.66 &	0.36 &	1.63 &	0.25 \\
$HR_{32}$	  &	1.57 &	0.48 &	1.53	 &0.44	 &	1.26 &	0.31 &	1.23 &	0.21 \\
\hline
\end{tabular}

Rest frame (145 GRBs; $T_{\rm l-s, i} = 2~$s, $25$ short, $120$ long): \\

\begin{tabular}{lcccccccc}
  \hline 
 & 		Short	& & & &				Long & & & \\

		& mean	& $\sigma_{\rm mean}$ & median & $\sigma_{\rm med}$ &	mean &	$\sigma_{\rm mean}$ & median & $\sigma_{\rm med}$  \\
  \hline 
$HR_{4321}$ 	&1.4	&1.05	&1.02	&0.7	&	1.21 &	0.57 &	1.11 &	0.38 \\
$HR_{43}$		&2.83	&1.50	&2.38	&1.01	&	2.63 &	0.84 &	2.51 &	0.57  \\
$HR_{41}$		&9.6	&11.08$^*$	&4.67	&3.9	&	6.68 &	4.72 &	5.28 &	2.72 \\
$HR_{31}$		&2.53	&1.70	&1.96	&1.11	&	2.25 &	0.92 &	2.10 &	0.62 \\
$HR_{21}$		&1.72	&0.57	&1.61	&0.38	&	1.68 &	0.34 &	1.66 &	0.23 \\
$HR_{32}$		&1.32	&0.49	&1.22	&0.33	&	1.28 &	0.29 &	1.26 &	0.19 \\
\hline
\end{tabular}

\scriptsize

 \begin{tabular}{l}
\end{tabular}

 \begin{tabular}{l}
Rest frame (145 GRBs; $T_{\rm l-s, i} = 2~{\rm s}/(1+\bar{z})=0.65~{\rm s}$, $8$ short, $137$ long): \\ 
 \end{tabular}

\begin{tabular}{lcccccccc}
  \hline 
	&	Short		&	&	& & Long & & &  \\
 & 		mean &	$\sigma_{\rm mean}$ &median	 &$\sigma_{\rm med}$	 &mean  &$\sigma_{\rm mean}$	 &median &$\sigma_{\rm med}$  \\
   \hline 
$HR_{4321}$ &	1.86 &	1.13 &	1.67	 &0.80	 &	1.21 &	0.63 &	1.06 &	0.41 \\
$HR_{43}$	  &	3.52 &	1.53 &	3.31	 &1.16	 &	2.61 &	0.93 &	2.45 &	0.60 \\
$HR_{41}$	 &	14.03 &	13.19 &	10.14 &	7.14	 &	6.78 &	5.53 &	4.94 &	2.67 \\
$HR_{31}$	  &	3.30 &	1.82 &	3.00	 &1.29	 &	2.24 &	1.02 &	2.02 &	0.68 \\
$HR_{21}$	  &	1.99 &	0.52 &	1.96	 &0.42	 &	1.67 &	0.37 &	1.63 &	0.25 \\
$HR_{32}$	  &	1.55 &	0.46 &	1.52	 &0.37	 &	1.27 &	0.32 &	1.23 &	0.21 \\	
\hline
\end{tabular}
\end{center}
\scriptsize
$^*$ In some cases, $\sigma_{\rm mean}$ exceeds the mean value due to large data scatter. $HR$s can have only non-negative values. \\
\end{table}

\section{Hardness ratio of Swift GRBs}
For hardness ratio analysis we use the following energy channels:
channel 1: 15-25~keV, channel 2: 25-50~keV, channel 3: 50-100~keV, and channel 4: 100-350 keV.
We define $HR_{\rm ij}$ as the ratio between fluence $F$ in channel ${\rm i}$ and ${\rm j}$, and $HR_{4321}={F_4}/(F_3+F_2+F_1)$.
We note that $HR_{321}$ in \cite{ref:kou} roughly corresponds to our $HR_{4321}$.

\begin{figure}[h]
\includegraphics[angle=-90,scale=0.235]{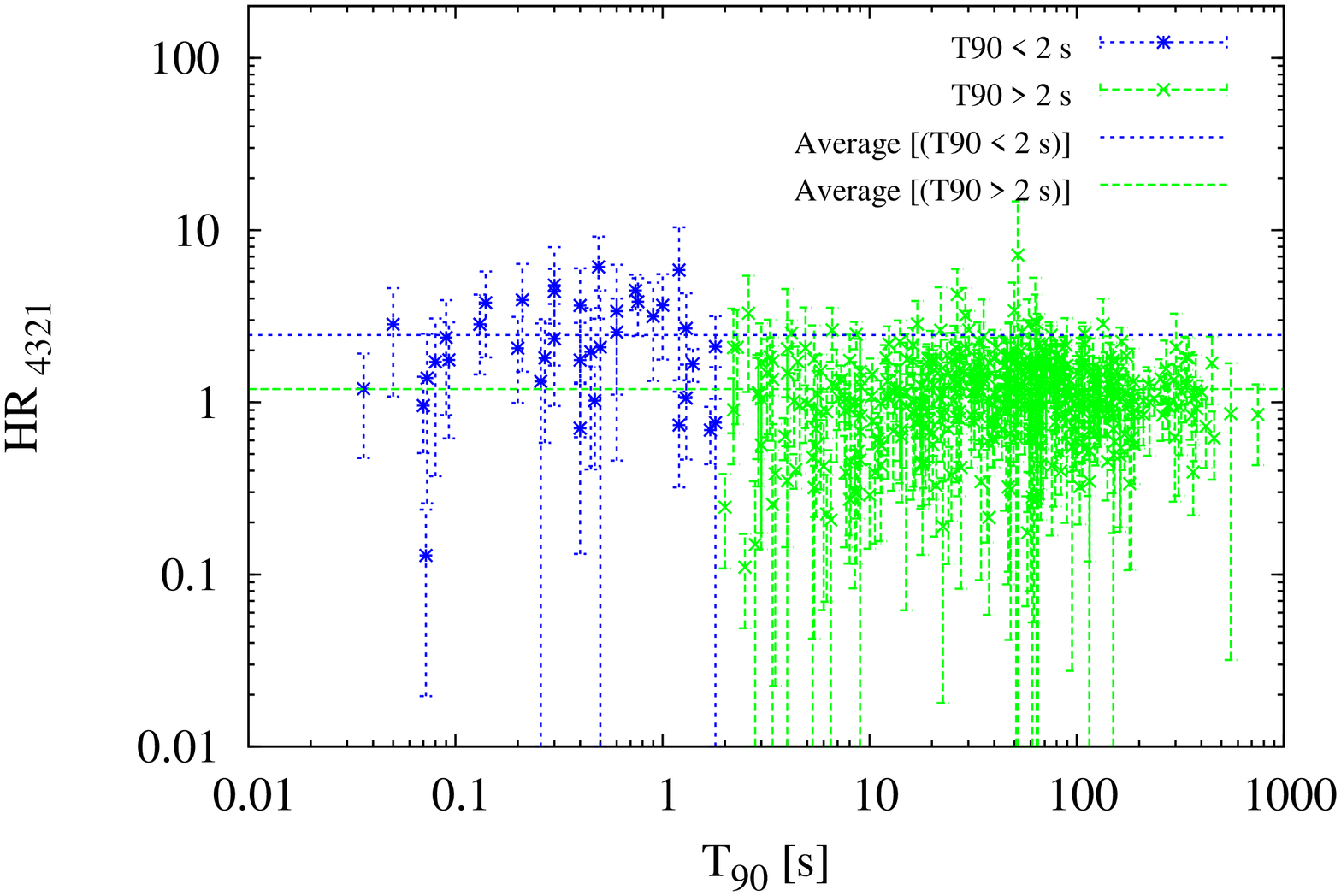}
\includegraphics[angle=-90,scale=0.235]{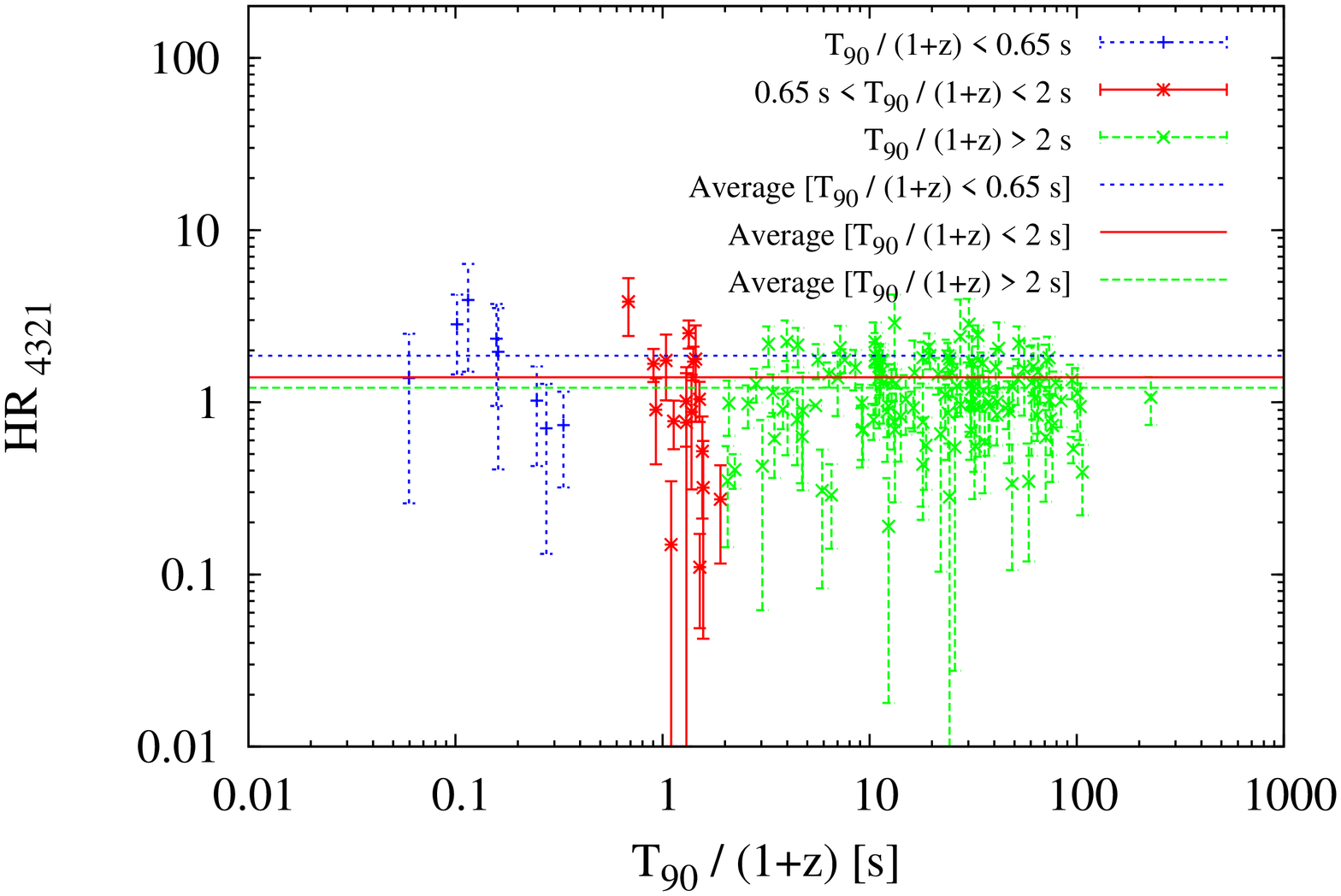}
\caption{$HR_{4321}$ vs burst duration in the observer frame (left) and in the rest frame (right). Horizontal lines represent average $HR_{4321}$ for short and long bursts.}
\label{fig02} 
\end{figure}

Results are presented in Table~\ref{HR} and Fig.~\ref{fig02} and depend on the channels used (${\rm i}$, ${\rm j}$).
First we analyze all GRBs and a subgroup of those with known redshift in the observer frame. We find the results for both groups to be consistent and the difference in $HR$s among short and long bursts not very pronounced.
Then we make a rough approximation by assuming that the hardness ratio of an individual burst is the same in the observer and in the rest frame (the transformation of fluences cancels out in the fluence ratio). In the transformation to the rest frame, only the burst duration $T_{90}$ is changed. If we take $T_{l-s, {\rm i}}=2~$s to separate bursts into short and long, we see that the difference in $HR$s between long and short burst decreases. In some cases the median value of $HR$s is even higher for long bursts than for short ones. On the other hand, by taking $T_{l-s, {\rm i}}=0.65~$s, we obtain results that are similar to the results in the observer frame for the same sample.

%

\end{document}